\documentclass[aps,prl,preprint]{revtex4-2}
\usepackage{graphicx}
\usepackage{color}
\usepackage{amsmath,amssymb}
\usepackage{comment}
\begin{document}

\title{Discrete Time Crystals in Noninteracting Dissipative Systems}


\author{Gourab Das$^{1,\dagger}$
Saptarshi Saha$^{2,\ddagger}$
and Rangeet Bhattacharyya$^{1,\star}$}

\affiliation{$^{1}$Department of Physical Sciences,
Indian Institute of Science Education and Research Kolkata, Mohanpur 741246, India\\
$^{2}$Institute for Physics and Astronomy (IFPA),
Technische Universit{\"a}t Berlin, EW 7-1, Hardenbergstr. 36, 10623 Berlin, Germany}

\begin{abstract}
Many-body quantum systems, under suitable conditions, exhibit time-translation symmetry breaking and settle in a discrete time crystalline (DTC) phase -- an out-of-equilibrium
quantum phase of matter.
The defining feature of DTC is a robust subharmonic
response. However, the DTC phase is fragile in the presence of environmental dissipation.
Here, we propose and exemplify a DTC phase in a noninteracting system that owes its
stability to environmental dissipation. The lifetime of this DTC is independent of initial
conditions and the size of the system, though it depends on the frequency of the external
driver. We experimentally demonstrate this realization of DTC using Nuclear Magnetic
Resonance spectroscopy. 
\end{abstract}

\maketitle

\section{Introduction}

Discrete time crystals (DTC) were proposed in periodically driven
quantum systems to exhibit discrete symmetry breaking in non-equilibrium systems
\cite{Shapere_Wilczek_2012, Wilczek_2012, Wilczek_2013, Bruno_2013,
Watanabe_Oshikawa_2015, Sacha_2015, Chandran_2016, Khemani_2016, Else_2016,
Keyserlingk_2016, Yao_2017}. Multiple groups experimentally realized the theoretical propositions over the last decade \cite{Chandran_2016, Khemani_2016, Else_2016, Keyserlingk_2016, Yao_2017,
Else_2017, Khemani_2017, Sacha_2015, Zhang_2017, Choi_2017}. These works defined the DTC phase as,
when a $T$-periodic drive, $H(t+T) = H(t)$, on a system produces a rigid reduced
periodicity for a certain physical observable $\hat{O}$, i.e. $\langle \hat{O} \rangle
(t+nT) = \langle \hat{O} \rangle (t)$ where $n$ is an integer and $n \geq 2$. The key feature of DTC is its robustness against the errors in the drive, promising its potential as a stabilizer against butterfly effects and heating \cite{Huang_2018}.

These recent theories employ non-ergodic many-body systems for resisting the
thermalization \cite{Khemani_2016, Else_2016, Keyserlingk_2016, Yao_2017}. The
interactions among the particles provide the robustness of DTC. As such, a stable
time-crystal has been shown to occur in the presence of the many-body localization (MBL)
regime with strong spatial disorder \cite{Else_2016, Yao_2017, Zhang_2017, Khemani_2016,
Else_2020, Khemani_2019, Choi_2017, Iemini_2024}. As well as DTCs have been engineered
with the help of bosonic self-trapping \cite{Sacha_2015, Sacha_2020, Sebastian_2012,
Russomanno_2017, Giergiel_2019, Matus_2019, Pizzi_2021}, gradient interaction
\cite{Yousefjani_2024}, Stark gradient fields \cite{Liu_2023, Kshetrimayum_2020},
confinement in domain-wall \cite{Collura_2022}, quantum scars \cite{Maskara_2021,
Deng_2023, Bull_2022, Huang_2022, Huang_2023}, and Floquet integrability
\cite{Huang_2018}. However, Choi \textit{et al.} demonstrated that MBL is not necessary by
an experiment on nitrogen-vacancy centers, where a time crystal (TC) formed regardless of the disorder in
the strongly interacting regime \cite{Choi_2017}. This experimental breakthrough gave
birth to a new kind of TCs, later named as \emph{clean time crystals} (cTC) \cite{Huang_2018}.
These cTCs exhibit integrability emerging through dynamics, unlike inheriting it from a static Hamiltonian, for the case of Floquet TCs \cite{Khemani_2016, Else_2016, Keyserlingk_2016, Yao_2017}.

These studies focus on the unitary dynamics in closed systems, but DTCs become fragile in
the presence of environmental dissipation \cite{Breuer_2002, Lazarides_2017, Riera_2020}. However, environmental dissipation can be a blessing in disguise for the
stability of a periodic observable, where the system releases excess energy pumped by a
periodic drive, with disturbance, to its environment via dissipation channels
\cite{Yousefjani_2024_b, Gong_2018, Gambetta_2019, Lazarides_2020}. These are called
dissipative DTC and have been observed experimentally \cite{Hans_2021, Taheri_2022}. In
this context, we have shown that small interacting quantum systems can give rise to
prethermal DTC, exploiting the emergence of a prethermal plateau during spin-locking
\cite{Saha_2024}

In this work, we show that (i) non-interacting quantum systems connected to a local environment can exhibit a robust subharmonic response like a DTC phase, (ii) the dissipation channels play crucial roles in stabilizing the DTC. 
Specifically, we show that having a short decoherence time and a long relaxation time
favors the formation of DTC in these systems. This phase is stabilized with only environmental dissipation,
without requiring interaction among the system's constituents, like space crystals forming
in the noninteracting systems \cite{Sacha_2020}.

\section{System}

We consider a single two-level system (TLS) having its local environment
acting as a heat bath. This TLS may be a magnetic dipole with the Zeeman Hamiltonian
$\mathcal{H}_\circ = \hbar \omega_\circ \sigma^z /2$. Hence, the evolution of the system
can be described by the Lindblad master equation as follows \cite{Breuer_2002},
\begin{equation}
\frac{d \rho}{dt} = -\frac{i}{\hbar} \left[ \mathcal{H}_\circ, \rho \right] + \sum_{j=1}^3 \left( L_j \rho L_j^\dagger - \frac{1}{2} \left\{ L_j^\dagger L_j, \rho \right\}\right)
\label{eq:1}
\end{equation}
where, $L_1 = \sqrt{\frac{1+M_\circ}{2T_1}} \sigma^+ $, $L_2 =
\sqrt{\frac{1-M_\circ}{2T_1}} \sigma^- $, $L_3 = \sqrt{\frac{1}{2T_\phi}} \sigma^z $ and
$M_\circ$ is the magnetization of the system. Hence, in the interaction representation 
(with respect to $\mathcal{H}_\circ$), the magnetization follows the following Bloch equations
\cite{Bloch_1946},
\begin{equation}
\frac{d M_x}{dt} = -\frac{M_x}{T_2}; \; \frac{d M_y}{dt} = -\frac{M_y}{T_2}; \; 
\frac{d M_z}{dt} = \frac{M_\circ - M_z}{T_1}
\label{eq:2}
\end{equation}
where, $\frac{1}{T_2} = \frac{1}{2T_1} + \frac{1}{T_\phi}$. Here, $T_1$ represents the
relaxation timescale, $T_{\phi}$ is the dephasing timescale, and $T_2$ is the decoherence timescale of the system, and they have
standard representation. Let us suppose that the environment has a longer relaxation
timescale compared to its decoherence timescale, i.e., $T_1 \gg T_2$.

\begin{figure}
\hspace*{-6mm}\raisebox{3.67cm}{(a)}\hspace*{1mm}
\includegraphics[width=0.7\linewidth]{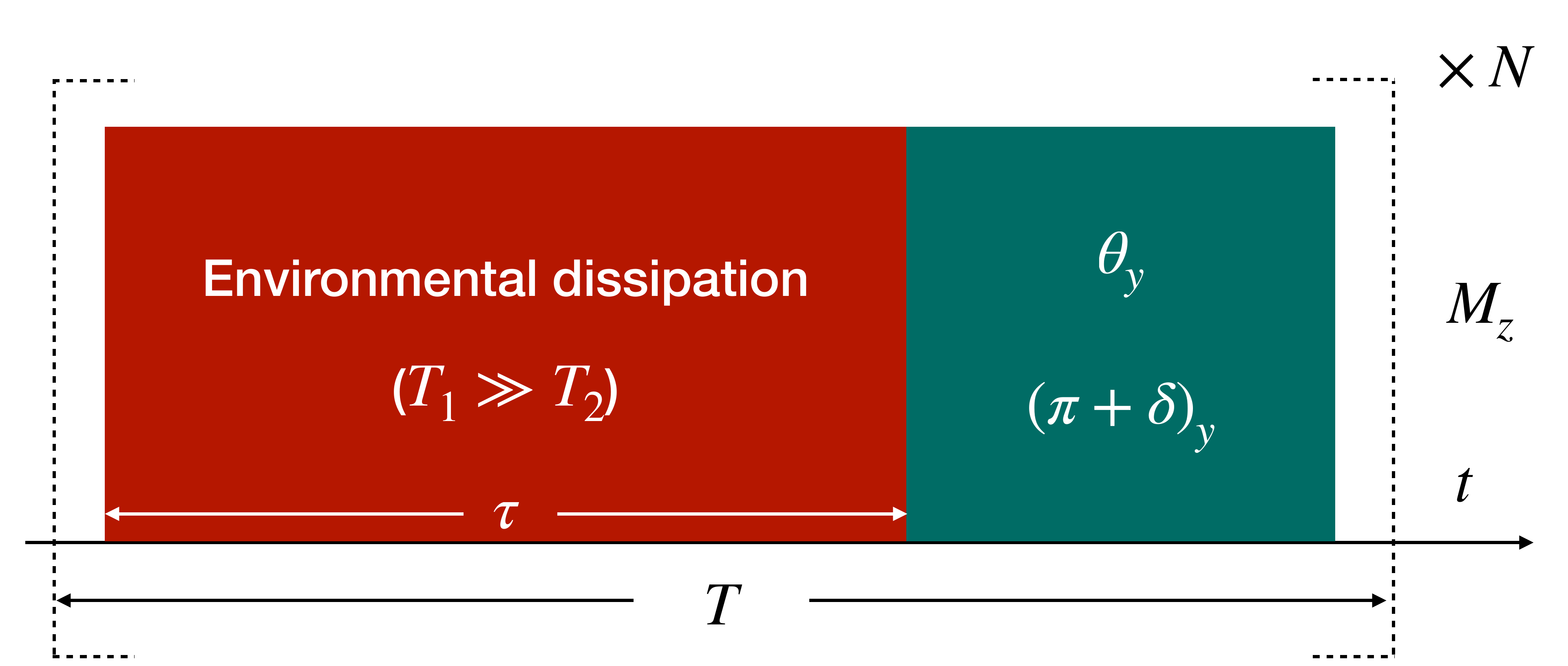}

\hspace*{-3mm}\raisebox{4.9cm}{(b)}
\includegraphics[width=0.48\linewidth]{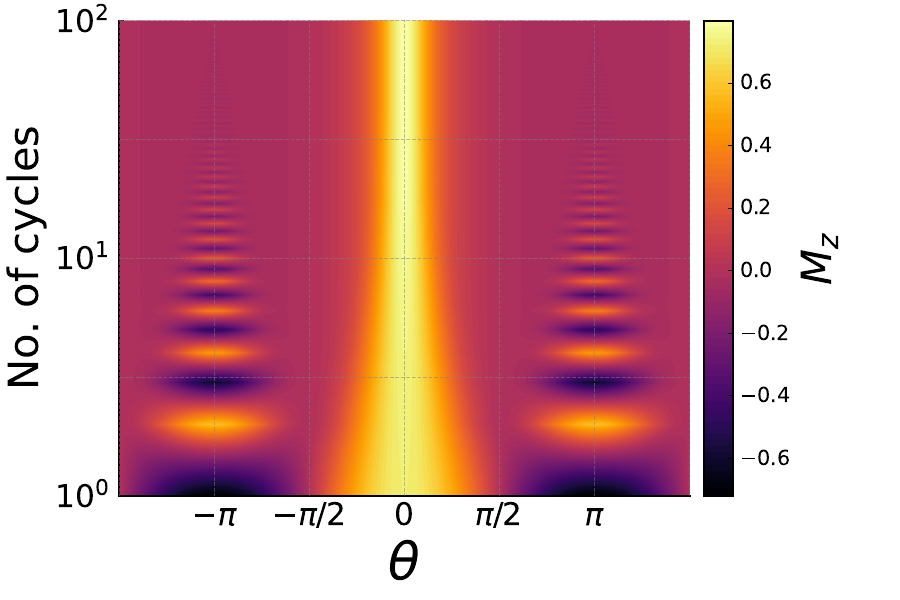}
\hspace*{-3mm}\raisebox{4.9cm}{(c)}
\includegraphics[width=0.48\linewidth]{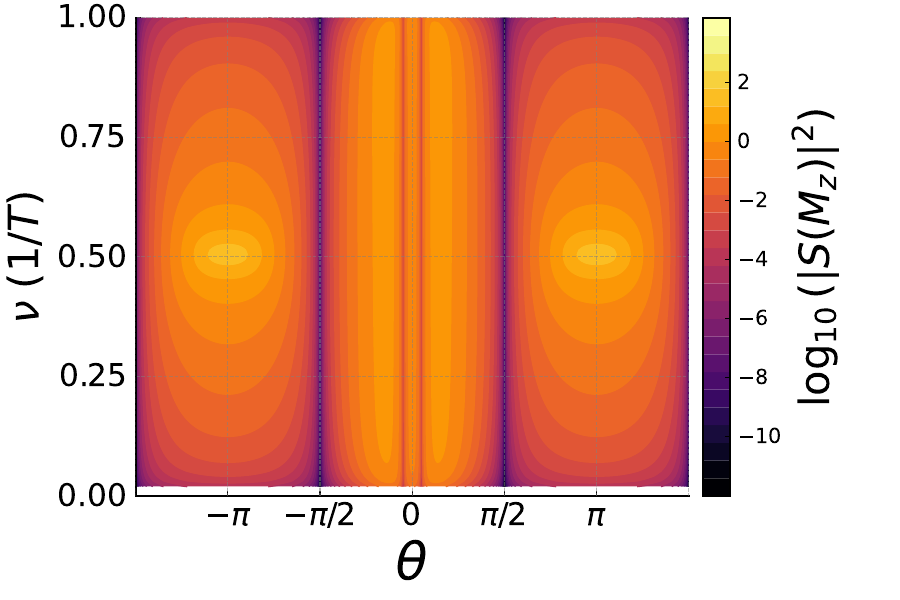}
\caption{(color online)
(a) Pulse sequence for environment-assisted DTC. Here, the system interacts with its
environment for a $\tau$ time period (red, deeper grey) and is followed by a $\theta$-pulse
about $y$-direction (green, lighter grey), and this sequence, having $T$ periodicity, is
repeated many times. As $T_1 \gg T_2$, relaxation of $M_z$ is slower than the decay of
$M_x$. Hence, the time delay $\tau$ acts similarly to a spin-locking pulse along
$z$-direction \cite{Choi_2017, Beatrez_2023}. Whilst, the $\theta$-pulse, when close to
$\pi$, provides the $2T$-periodicity for $M_z$.  (b) The existence of DTC depends on the
$\theta$-pulse. Here $M_z$ oscillates between positive and negative values in alternate
cycles when $\theta$ is close to $\pm \pi$. (c) Spectrum of $M_z$ in the frequency
domain. The Fourier spectrum shows peaks around $\nu = 0.5$ frequency, confirming the
$2T$-periodicity of $M_z$ in the same regions of $\theta$.  The parameters used to
generate the plots of (b) and (c) are $T_1 = 100 T_2$, $\tau = 10 T_2$, $M_\circ = 0.8$,
and $M_z(0) = -0.9M_\circ$.
}
\label{fig:1}
\end{figure}

Next, a $\theta = \pi + \delta$ rotation is applied about the $y$-direction, as shown in
Fig. \ref{fig:1}. The corresponding Hamiltonian, in the interaction picture of
$\mathcal{H}_\circ$, is $H_y = \hbar \omega_1 \sigma^y/2$, where $\omega_1 (T-\tau) = \pi
+ \delta$. Usually, a strong pulse is provided for the $\theta$-rotation to keep the pulse
duration small, hence the dissipation during the excitation pulse is ignored
\cite{Beatrez_2023}. The corresponding dynamics is governed by the von Neumann-Liouville
equation, as follows \cite{Breuer_2002},
\begin{equation}
\frac{d \rho}{dt} = -\frac{i}{\hbar} \left[ H_y, \rho \right] \quad \text{for } \tau < t \leq T 
\label{eq:3}
\end{equation}

To obtain $2T$-periodicity the mentioned sequence, in Fig. \ref{fig:1}(a), is repeated $N$
times $\left( N \in \mathcal{I} \right)$. Hence, the final density matrix, in the
Liouville space, after $N$-cycle can be written as,
\begin{equation}
\hat{\rho}(NT) = \left[ e^{\hat{\mathcal{L}}_y (T-\tau)} e^{\hat{\mathcal{L}}_\tau \tau} \right]^N \hat{\rho}(0)
\label{eq:4}
\end{equation}
Here, $\hat{\mathcal{L}}_y$ is the Liouvillian refers to the rotation about $y$ direction,
from Eq. \ref{eq:3}, and $\hat{\mathcal{L}}_\tau$ is the same corresponding to system's
interaction with its environment during the $\tau$-time delay, from Eq. \ref{eq:2}. 

\section{Emergence of the DTC phase}

The system's dynamics can be tracked easily in the
parts of the pulse sequence in Fig. \ref{fig:1}(a). During the time delay $\tau$, the
system follows the Bloch equations \cite{Bloch_1946}, as given in Eq. \ref{eq:2}. For the
initial magnetization $\mathbf{M} = (M_x^0, M_y^0, M_z^0)$, the solutions are,
\begin{equation}
\begin{gathered}
M_x(t) = M_x^0 e^{-\frac{t}{T_2}} \equiv M_x^\tau \left(M_x^0, t \right)\\
M_y(t) = M_y^0 e^{-\frac{t}{T_2}} \equiv M_y^\tau \left(M_y^0, t \right)\\
M_z(t) = M_\circ \left(1-e^{-\frac{t}{T_1}} \right) + M_z^0 e^{-\frac{t}{T_1}} \equiv M_z^\tau \left(M_z^0, t \right)
\end{gathered}
\label{eq:5}
\end{equation}
Therefore, after the $\tau$ time delay $M_{x/y}^\tau (M_{x/y}^0,\tau) \approx 0$ and
$M_z^\tau(M_z^0, \tau) \approx M_z^0$, if $T_2 < \tau \ll T_1$. Hence, $M_z$ relaxes
slower than the decay of other magnetization components; in turn, a time delay of $\tau$, with this timescale separation, acts similarly to the spin-locking pulse along the $z$-direction
\cite{Choi_2017, Beatrez_2023, saha2023cascaded, chakrabarti2023emergence, Saha_2024}, i.e. required for the rigidity of DTC phase
\cite{Huang_2018}. Therefore, we can use a pulse sequence, similar to Beatrez \textit{et
al.} \cite{Beatrez_2023}, where the system interacts with the environment for $\tau$ time
duration, instead of the spin-locking pulse, as shown in Fig. \ref{fig:1}(a).

As the $\theta$ rotation around the $y$-direction occurs for a short duration, the dissipation
can be neglected during this period. Therefore, the dynamics are adequately described by
the first-order process.

\subsubsection{Solution for $\theta = \pi$}

For a perfect $\pi$ pulse, with initial condition
$\mathbf{M} = (0, 0, M_z^0)$ we get (see Appendix \ref{A:1} for more details),
\begin{align}
&M_z(T) = - M_z^\tau \left(M_z^0, \tau \right) \nonumber \\
&M_z(2T) = - M_z^\tau \left(- M_z^\tau \left(M_z^0, \tau \right), \tau \right)
\label{eq:6}
\end{align} 

Here, all other magnetization components vanish. As $\tau \ll T_1$, $M_z$ reverses its
sign after time $T$, returning near the initial value after $2T$ time, i.e., after
the second $\pi$-pulse. Hence, the perfect $\pi$ pulse provides $2T$-periodicity producing
DTC phase. However, the DTC vanishes in the timescale of $T_1$, as $ M_z$ relaxes with
$T_1$ timescale. As the system consists of noninteracting particles, the lifetime of this DTC phase is independent of particle numbers, unlike previously reported DTCs \cite{Huang_2018}.

\begin{figure}
\hspace*{-3mm}\raisebox{5.0cm}{(a)}\hspace*{1mm}
\includegraphics[width=0.5\linewidth]{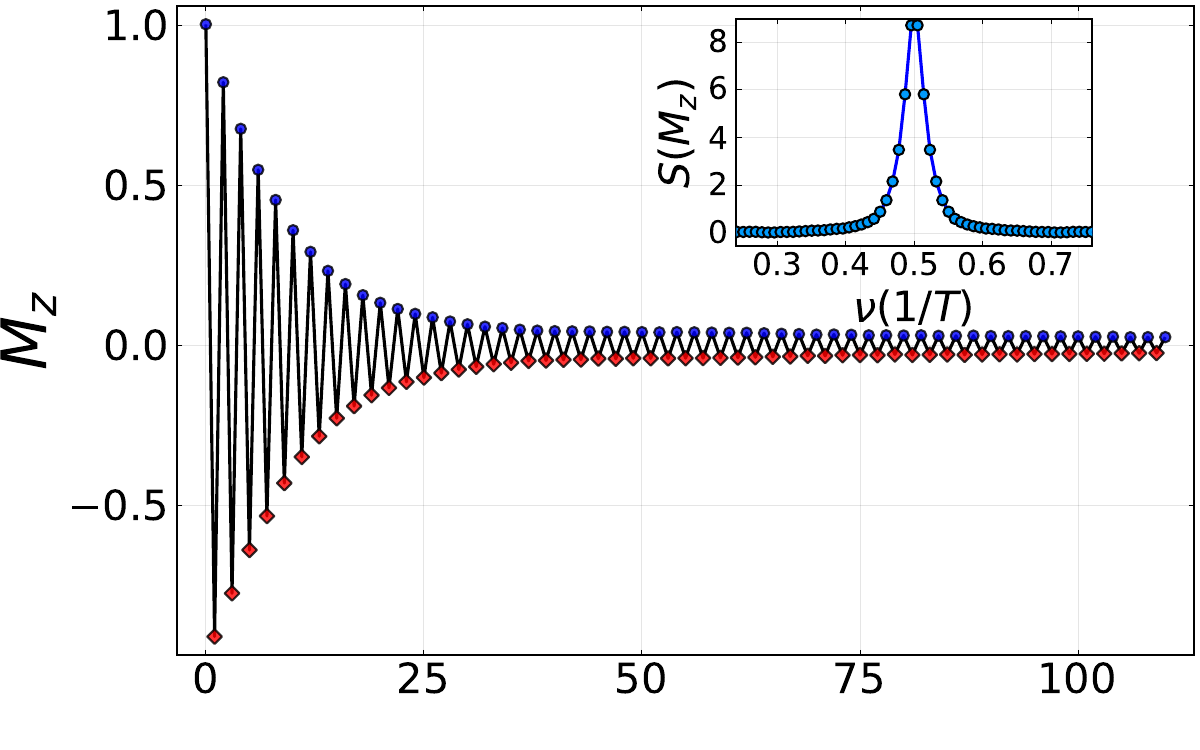}\\
\hspace*{-3mm}\raisebox{5.0cm}{(b)}\hspace*{1mm}
\includegraphics[width=0.5\linewidth]{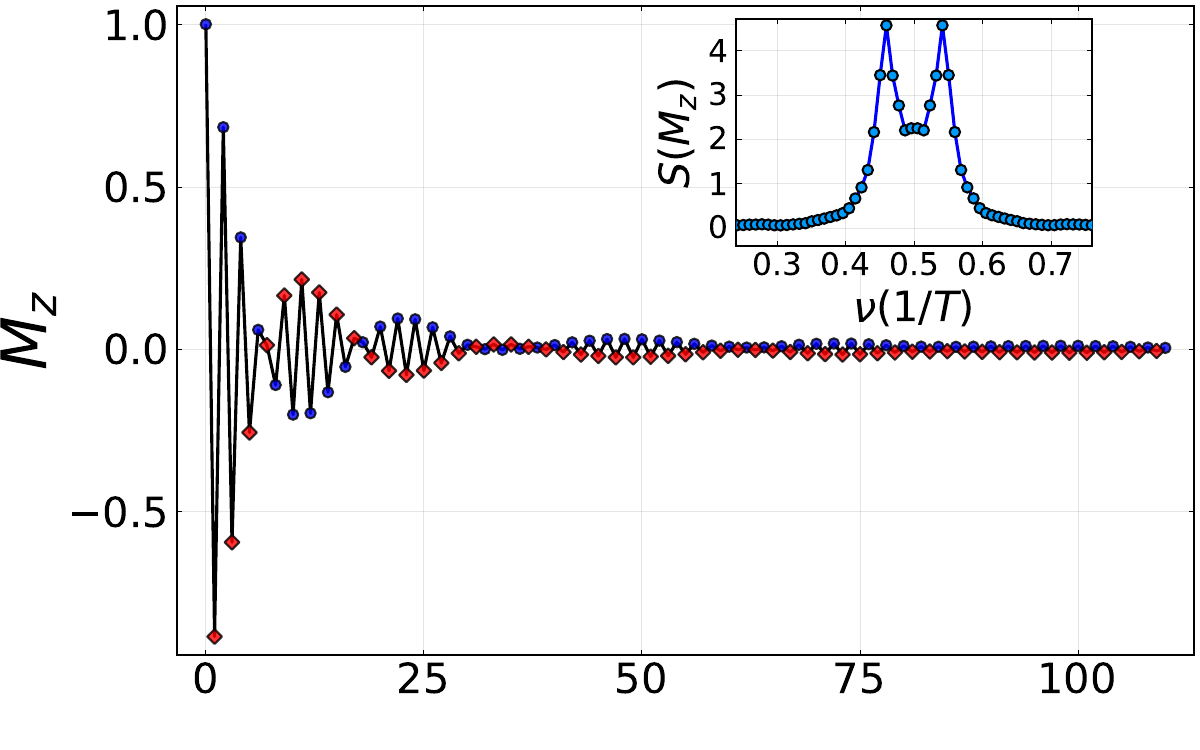}\\
\hspace*{-3mm}\raisebox{5.0cm}{(c)}\hspace*{1mm}
\includegraphics[width=0.5\linewidth]{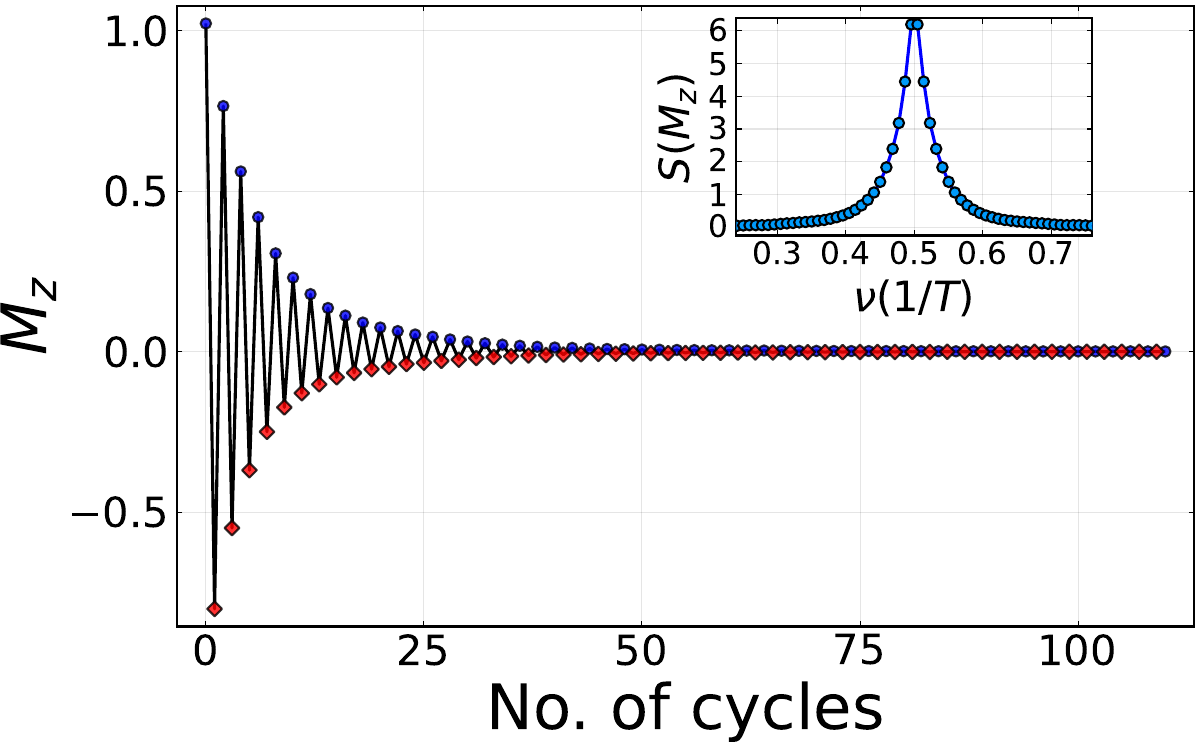}
\caption{(color online)
Plots of experimentally determined values of $M_z$ versus the number of cycles are shown in
(a), (b), (c) and their corresponding Fourier transform, $S(M_z)$, versus frequency
($\nu$) are shown in the insets. (a) Perfectly calibrated $\pi$ pulses are used to get
the DTC phase, which produces a sharp peak at $0.5$ in the frequency domain. Here $\tau =
25ms$.  (b) In this case the pulses were with error of $\delta = 0.0674 \pi$, with
$\tau = 25ms$, kills the DTC phase to produce a decaying beat pattern. This corresponds to
peaks at $0.5 \left( 1 \pm \delta \right)$ in the frequency domain. (c) Here, the time
delay has been increased, $\tau = 0.2s$, for the same imperfect pulse of (b) to get back
DTC, showing the rigidity of such a phase. However, the lifetime of DTC has reduced
significantly.
}
\label{fig:2}
\end{figure}

\subsubsection{Solution for $\theta = \pi +\delta$ with $\delta/\pi \to 0$}

Here, we check the
rigidity of DTC phase by applying a perturbation ($\delta$) in the $\pi$ rotation, i.e.
$\theta = \pi + \delta$, and we get,
\begin{widetext}
\begin{equation}
\begin{gathered}
M_x(T) = - M_z^\tau \left(M_z^0, \tau \right) \delta; \quad
M_x(2T) = - M_x^\tau \left(- M_z^\tau \left(M_z^0, \tau \right) \delta, \tau \right) - M_z^\tau \left(- M_z^\tau \left(M_z^0, \tau \right), \tau \right) \delta\\
M_z(T) = - M_z^\tau \left(M_z^0, \tau \right) ; \quad
M_z(2T) = - M_z^\tau \left(- M_z^\tau \left(M_z^0, \tau \right), \tau \right) + M_x^\tau \left(- M_z^\tau \left(M_z^0, \tau \right) \delta, \tau \right)\delta
\end{gathered}
\label{eq:7}
\end{equation}
\end{widetext}

\begin{figure}
\hspace*{-3mm}\raisebox{4.7cm}{(a)}\hspace*{1mm}
\includegraphics[width=0.47\linewidth]{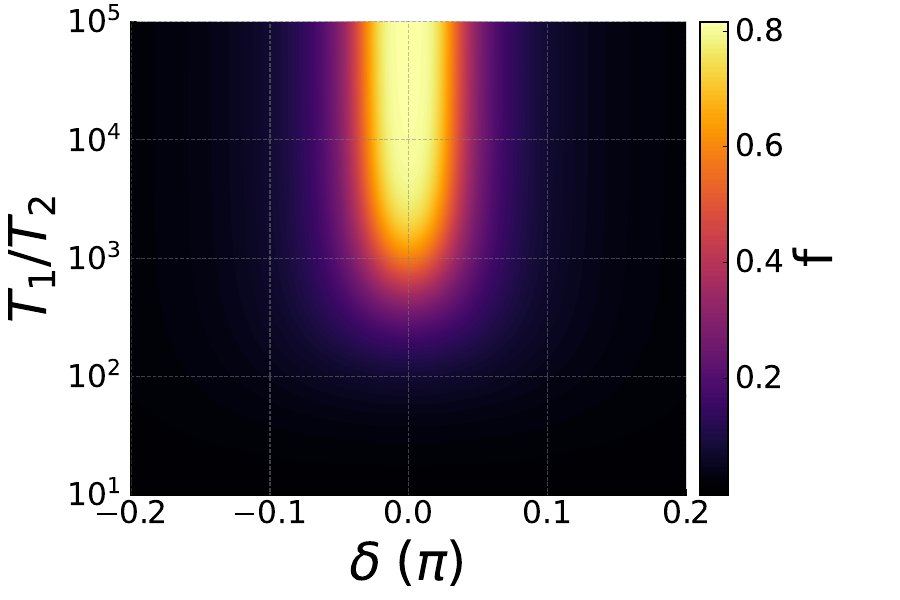}
\hspace*{-3mm}\raisebox{4.7cm}{(b)}\hspace*{1mm}
\includegraphics[width=0.47\linewidth]{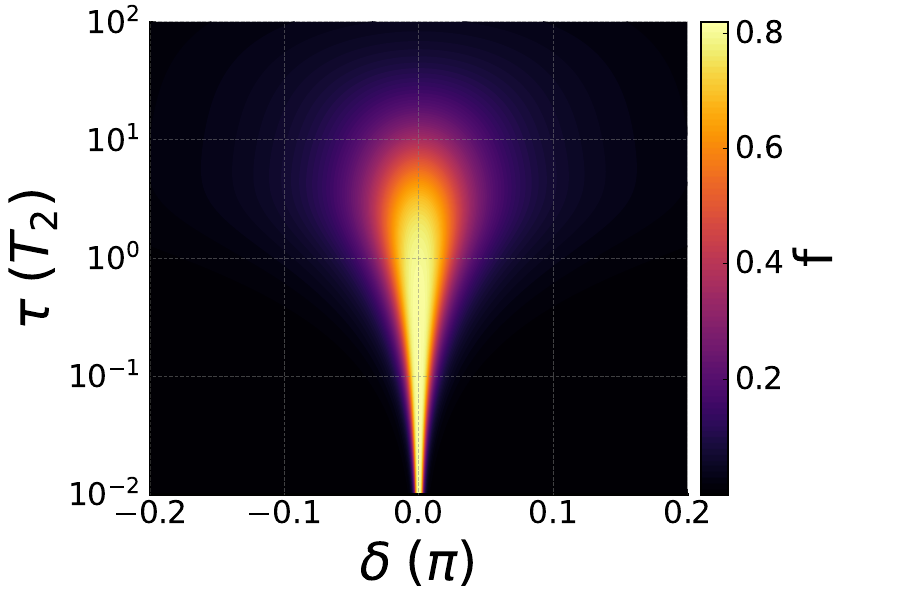}\\
\hspace*{-3mm}\raisebox{7.0cm}{(c)}\hspace*{1mm}
\includegraphics[width=0.7\linewidth]{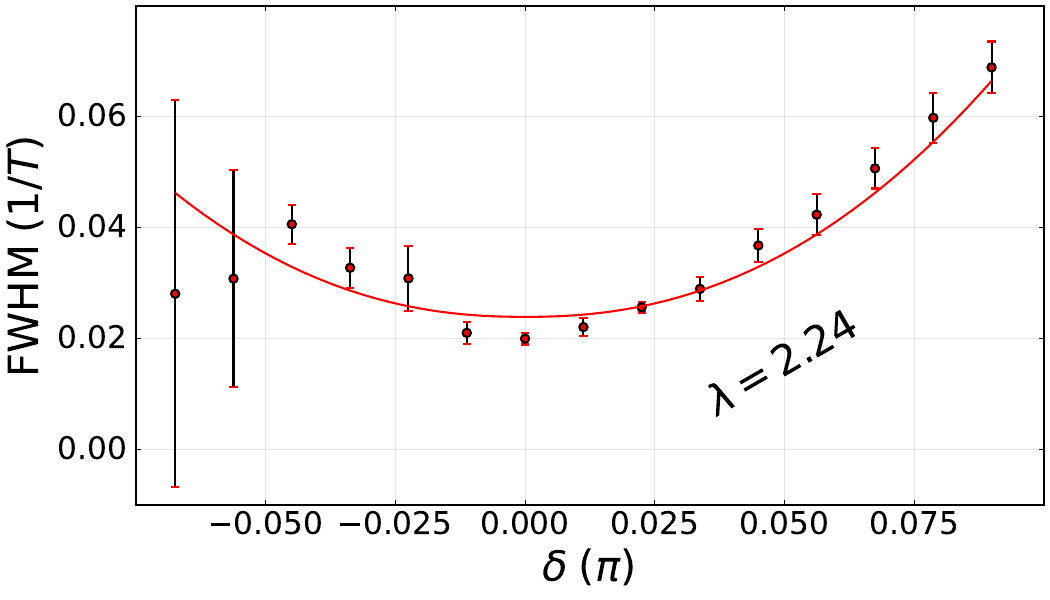}
\caption{(color online)
(a) Crystalline fraction (f), defined by Choi \textit{et al.} \cite{Choi_2017},
dependence on the perturbation ($\delta$) in the $\pi$ rotation about $y$-direction and
the ratio of the timescales $T_1$ and $T_2$. The Crystalline fraction, f, acts as a
measure for the amount of DTC phase present in the system. To have DTC phase $\delta$
needs to be small, i.e. we need near perfect $\pi$ pulses and $T_2 \ll T_1$, as it follows
from Eq. \ref{eq:7}. We have used $\tau = 5 T_2$ for the plot. (b) Crystalline fraction's
dependence on $\delta$ and time interval $\tau$ (in the units of $T_2$). It shows $\delta$ needs to be small and
$T_2 < \tau \ll T_1$ for having DTC phase, following Eq. \ref{eq:7}. Here, $T_1 = 1000
T_2$ is considered for the plot. The other parameters used to generate the plots, of (a)
and (b), are $M_\circ = 0.8$, and $M_z(0) = -0.9M_\circ$.  (c) Plot of the FWHM of the
Fourier spectrum of $M_z$, i.e. $S(M_z)$, versus the error ($\delta$) in $\pi$ pulse
(experimental data for $\tau = 25ms$). The data is fitted with $f(\delta) = a \delta ^\lambda + b$ for $\lambda = 2.24$. Hence, FWHM varies as $\delta^2$; therefore,
the lifetime of the DTC phase decreases at the same rate. 
}
\label{fig:3}
\end{figure}

Here, the period doubling of $M_z$ is absent in the Eq. \ref{eq:7} due to the $M_x^\tau$
term. DTC phase can be retrieved if $\delta \to 0$ or $T_2 < \tau$. However, we need that
$\tau \ll T_1$, otherwise $M_z$ will relax back near to $M_\circ$ preventing the existence
of DTC, as shown in Fig. \ref{fig:2} and \ref{fig:3}(a,b). The lifetime, which is
inversely proportional to the full width at half maximum (FWHM) of $M_z$ in the frequency
domain \cite{Saha_2024}, of DTC phase vanishes as $\delta^2$ due to the error term in
$M_z$ varies at the same rate (Eq. \ref{eq:7}), as shown in Fig. \ref{fig:3}(c). We name this environment-assisted DTC as EDTC.

\section{Experiment}
\begin{figure}
\includegraphics[width=0.7\linewidth]{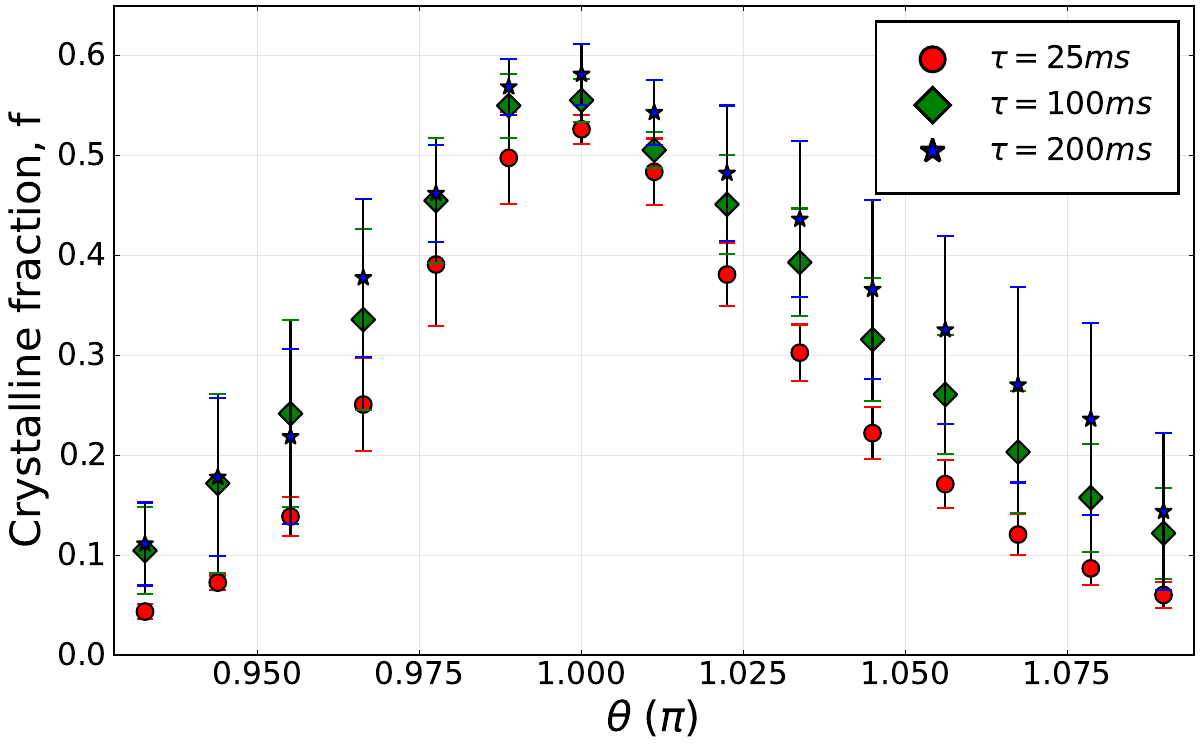}
\caption{(color online)
The variation of Crystalline fraction, f, with $\theta$ pulse is plotted, with different time delays $\tau$. It shows that we need a pulse very close to $\pi$, for a short $\tau$, to get the EDTC phase. Whilst EDTC persists for imperfect pulses when a longer $\tau$ is applied,
showing the rigidity of the phase.
}
\label{fig:4}
\end{figure}

We use Nuclear Magnetic Resonance, an ensemble spectroscopy, to experimentally demonstrate our proposed scheme of realizing EDTC. As the sample, we use $99.9\% \; D_2 O$, which is
effectively a collection of $HDO$
molecules to show the existence of such a phase.
The nucleus of the Hydrogen atom, a lone proton, in the $HDO$ molecule, serves as the spin-1/2 particle.
We used a Bruker Avance III $500$ MHz spectrometer to perform all the experiments, and the
ambient temperature during the measurements was $25^{\circ}$C.
We determined the relaxation times of protons in $HDO$ as $T_1 \simeq 7.57s, \; T_2^* \sim 0.6s$. We used a uniform 16.7 kHz RF drive amplitude to apply various pulses in all the experiments. 
The variation of
crystalline fraction, f, with respect to the $\theta$ pulse along the $y$-direction for three
sets of time delay, $\tau$ ($25ms, 100ms, 200ms$), is shown in Fig. \ref{fig:4}. It shows
that we need near perfect $\pi$ pulse to have an EDTC phase, however a longer $\tau$ can
mitigate higher errors, $\delta$, to preserve EDTC, thus validating the rigidity of such a phase.

\section{Discussion}
\begin{figure}
\includegraphics[width=0.7\linewidth]{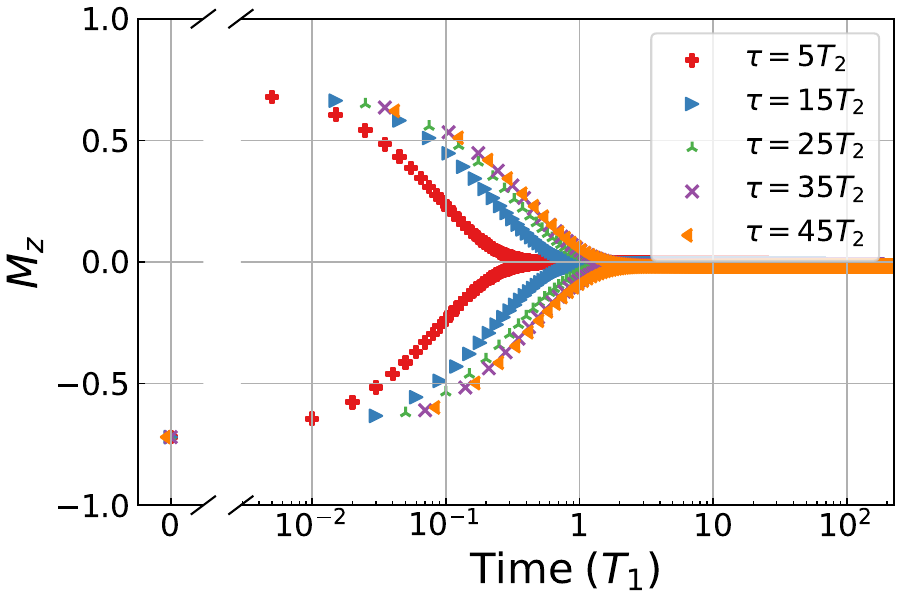}
\caption{(color online)
The variation of $M_z$ with time (in the units of $T_1$) is shown for different time delays $\tau$'s. This shows the lifetime of this DTC increases with $\tau$. Hence, the lifetime decreases with increasing frequency of the pulse. The parameters used to generate
the plot are $T_1 = 1000 T_2$, $\delta = 0.1 \pi$, $M_z(0) = -0.9M_\circ$, and $M_\circ = 0.8$.
}
\label{fig:5}
\end{figure}

We note the scheme relies on isolated TLS with its environment. As
such, the experiment does not depend on the number of TLS chosen as long as the number is
thermodynamically large. The FWHM of $S(M_z)$ does not depend on the initial
magnetization, which is related to the initial state's temperature via the Boltzmann
distribution. As such, the lifetime of the EDTC, like Floquet DTC phase, does not depend on the initial condition \cite{Khemani_2016, Keyserlingk_2016, Keyserlingk_2016_b,
Else_2016, Yao_2017, Zhang_2017}. However, Fig. \ref{fig:5} show that it depends on
the time delay $\tau$, which is inversely proportional to the frequency of the $\theta$
pulse, when an erroneous pulse is provided, i.e., a property of prethermal TCs, though the lifetime is independent of the frequency in case of perfect pulses, as the dynamics is only governed by $T_1$ and $T_2$ dynamics
\cite{Machado_2020, Saha_2024}. Hence, the EDTC phase is not a  Floquet or prethermal DTC. We note the decoherence dissipator governs the system's dynamics during the time delay, $\tau$. Hence, unlike the other DTCs, the EDTC phase lacks the Hamiltonian description and instead has a Liouvillian description.

\section*{Conclusions}

To conclude, we show that the DTC phase, if defined through robust subharmonic response of a quantum system, having many-body interactions is not a strict necessity; the environment
can provide robustness to dissipative systems. This is in line with our previous work
demonstrating that a prethermal plateau can emerge in small dissipative systems too
\cite{Saha_2024}. We show that an open quantum system, with the decoherence timescale much shorter compared to that of the relaxation ($T_2 \ll T_1$), can be manipulated to have an environment-assisted DTC phase. In fact, this EDTC
phase does not require interactions among the system's constituents; hence, the lifetime
of such a phase is independent of system size. This novel phase is neither a Floquet nor a prethermal DTC; also, it lacks a Hamiltonian description as a dissipator governs the dynamics during the $\tau$-delay. We demonstrate the existence of such an EDTC phase using nuclear magnetic resonance spectroscopy.
 

\textit{Acknowledgments - }GD gratefully acknowledges the Council of Scientific \&
Industrial Research (CSIR), India, for a research fellowship (File no:
09/921(0327)/2020-EMR-I). The authors thank Arpan Chatterjee, Sarfraj Fency, Shubhamay Panja, and Arkadeep Mitra for their insightful comments. 

\bigskip
\noindent
$^{\dagger}$gd20rs094@iiserkol.ac.in\\
$^{\ddagger}$s.saha@tu-berlin.de\\
$^{\star}$rangeet@iiserkol.ac.in

\bibliographystyle{unsrt}
\bibliography{ref}

\appendix
\setcounter{secnumdepth}{2} 

\section{Dynamics to the DTC phase} \label{A:1}

As the $\theta$ rotation around $y$-direction occurs for a short duration, the dissipation
can be neglected during this period. Therefore, the dynamics are adequately described by
the first-order process, with the initial magnetization $\mathbf{M} = (m_x^0, m_y^0,
m_z^0)$ the solution for Eq. \ref{eq:3} is given as,
\begin{equation}
\begin{gathered}
M_x(t) = m_x^0 \cos \theta + m_z^0 \sin \theta \equiv M_x^\text{rot} \left(m_x^0, m_z^0, \theta \right)\\
M_y(t) = m_y^0 \equiv M_y^\text{rot} \left(m_y^0, \theta \right)\\
M_z(t) = m_z^0 \cos \theta - m_x^0 \sin \theta \equiv M_z^\text{rot} \left(m_x^0, m_z^0, \theta \right)
\end{gathered}
\label{eq:A.1}
\end{equation}
Therefore, $M_x^\text{rot} \left(m_x^0, m_z^0, \theta \right) \approx -m_x^0$,
$M_y^\text{rot} \left(m_y^0, \theta \right) = m_y^0$, and $M_z^\text{rot} \left(m_x^0,
m_z^0, \theta \right) \approx -m_z^0$ after the $\theta$-pulse for $\theta \approx
\pm\pi$.

To study the emergence of the DTC phase, we provide the analytical solution of $M_z(t)$ up
to $2T$ time period by solving Eq. \ref{eq:4}. For initial condition $\mathbf{M} = (0, 0,
M_z^0)$, the magnetization at time $\tau$ is given as,
\begin{equation}
\begin{gathered}
M_x(\tau) = 0; \quad
M_z(\tau) = M_z^\tau \left(M_z^0, \tau \right)
\end{gathered}
\label{eq:A.2}
\end{equation}
After the $\theta$ rotation about $y$-direction, the magnetization is given by,
\begin{equation}
\begin{gathered}
M_x(T) = M_z^\tau \left(M_z^0, \tau \right) \sin \theta ; \quad
M_z(T) = M_z^\tau \left(M_z^0, \tau \right) \cos \theta
\end{gathered}
\label{eq:A.3}
\end{equation}
Similarly, after a time $T+\tau$, the magnetization is,
\begin{equation}
\begin{gathered}
M_x(T+\tau) = M_x^\tau \left( M_z^\tau \left(M_z^0, \tau \right) \sin \theta, \tau \right) \\
M_z(T+\tau) = M_z^\tau \left( M_z^\tau \left(M_z^0, \tau \right) \cos \theta, \tau \right)
\end{gathered}
\label{eq:A.4}
\end{equation}
Finally, after the $2T$ time, the magnetization becomes,
\begin{widetext}
\begin{equation}
\begin{gathered}
M_x(2T) = M_x^\tau \left( M_z^\tau \left(M_z^0, \tau \right) \sin \theta, \tau \right) \cos \theta + M_z^\tau \left( M_z^\tau \left(M_z^0, \tau \right) \cos \theta, \tau \right) \sin \theta \\
M_z(2T) = M_z^\tau \left( M_z^\tau \left(M_z^0, \tau \right) \cos \theta, \tau \right) \cos \theta - M_x^\tau \left( M_z^\tau \left(M_z^0, \tau \right) \sin \theta, \tau \right) \sin \theta
\end{gathered}
\label{eq:A.5}
\end{equation}
\end{widetext}
Here, $M_y$ does not evolve, as $y$ is the symmetry axis.

\end{document}